\documentclass[journal,transmag]{IEEEtran}
 \pdfoutput=1
 \usepackage{dcolumn}% Align table columns on decimal point 
 \usepackage{graphicx,color}
 \usepackage[cmex10]{amsmath} 
 \usepackage{amssymb} % for \lesssim and freinds
 \usepackage{array}
 \usepackage{cite}

 \PassOptionsToPackage{caption=false}{subfig}
 \usepackage{subfig}
 \usepackage{url}

\begin{document}

\author{\IEEEauthorblockN{P.~I.~Bunyk,
 E.~Hoskinson,
 M.~W.~Johnson,
 E.~Tolkacheva,
 F.~Altomare,
 A.~J.~Berkley,\\
 R.~Harris,
 J.~P.~Hilton,
 T.~Lanting,
 J.~Whittaker}
\thanks{The authors are with D-Wave Systems Inc., 3033 Beta Avenue, Burnaby BC Canada V5G 4M9}
\thanks{Corresponding author: P.~I.~Bunyk (email: pbunyk@dwavesys.com)}
}

% The following two lines redefine the subscript operator _
% so that it always uses \mathrm. Also, the \ensuremath means
% it can be used in regular text.
% For original subscript behaviour, use \sb instead of _
\catcode`_=\active
\newcommand_[1]{\ensuremath{\sb{\mathrm{#1}}}}

\newcommand{\fdac}{$\Phi$-DAC{}}

%\IEEEtitleabstractindextext{%

%}

\title{Architectural considerations in the design of a superconducting
  quantum annealing processor}

\maketitle

\begin{IEEEkeywords}
  Superconducting integrated circuits, Quantum computing,
  Computational physics
\end{IEEEkeywords}

\begin{abstract}
  We have developed a quantum annealing processor, based on an array
  of tunably coupled rf-SQUID flux qubits, fabricated in a
  superconducting integrated circuit
  process\cite{harris2010b}. Implementing this type of processor at a
  scale of 512 qubits and 1472 programmable inter-qubit couplers and
  operating at $\mathbf{\sim 20\, mK}$ has required attention to a
  number of considerations that one may ignore at the smaller scale of
  a few dozen or so devices. Here we discuss some of these
  considerations, and the delicate balance necessary for the
  construction of a practical processor that respects the demanding
  physical requirements imposed by a quantum algorithm. In particular
  we will review some of the design trade-offs at play in the
  floor-planning of the physical layout, driven by the desire to have
  an algorithmically useful set of inter-qubit couplers, and the
  simultaneous need to embed programmable control circuitry into the
  processor fabric.  In this context we have developed a new ultra-low
  power embedded superconducting digital-to-analog flux converters
  (DACs) used to program the processor with zero static power
  dissipation, optimized to achieve maximum flux storage density per
  unit area. The 512 single-stage, 3520 two-stage, and 512 three-stage
  flux-DACs are controlled with an XYZ addressing scheme requiring 56
  wires.  Our estimate of on-chip dissipated energy for worst-case
  reprogramming of the whole processor is $\mathbf{\sim 65\, fJ}$.
  Several chips based on this architecture have been fabricated and
  operated successfully at our facility, as well as two outside
  facilities (see for example \cite{natureNewsGoogle}).
\end{abstract}

%\IEEEdisplaynontitleabstractindextext

\tableofcontents
  % Prominent among these considerations is the
  % necessity of interacting with so many devices in a way that does not
  % require the delivery of thousands of separate electrical signals to
  % a single, cryogenically operated processor chip.

\section{Introduction}

\IEEEPARstart{P}{roposed} implementations of quantum computers capable
of solving problems at a useful scale generally involve at least many
thousands of qubits.  Whether the algorithm envisioned is based on the
quantum circuit model, or on an adiabatic method, there are a number
of physical requirements that constrain the design of a large-scale
manufactured quantum device. The device architecture must facilitate
precise individual qubit control, computationally interesting
interaction between qubits, and high fidelity readout of qubit
state. Of particular importance are the practical constraints that
arise in acheiving these design goals while maintaining the carefully
engineered environment required to implement a quantum algorithm.

One of the advantages of an approach based on superconducting qubits
is that a largely compatible classical electronics technology is known
and available in the guise of single-flux-quanta based circuit
architectures.  The authors have previously presented the
architecture, design, and operation of an SFQ based system for
controlling\cite{rainierPMM} and reading out\cite{rainierREADOUT} a
quantum annealing processor based on flux
qubits\cite{harris2010b,natureQAMS} at the core of D-Wave
One\texttrademark\ system. The authors, as well as a number of other
researchers, have gained some experience using this first generation
processor \cite{QA100,Pudenz2013,Gaitan2013,NIPS2009}.  This
experience has informed and guided the design of a second generation
quantum annealing processor, the D-Wave Two\texttrademark\ system.  In
this paper we provide an overview of this processor's architecture and
discuss some of the considerations and trade-offs involved in its
design.

At the root of the design problem is the operation of a quantum
annealing processor based on superconducting flux qubits. Many of the
processor building blocks are described by Harris, {\em et
  al.}\cite{harris2010b,harris2010a}.  A number of the control
terminals (biases) required by the qubits and inter-qubit couplers
are discussed by Johnson {\em et al.}\cite{rainierPMM}, and these are
largely unchanged. The implementation of the control circuitry, the
cross section of the fabrication process, and the number of devices
used, are among the main differences between the two generations of
D-Wave processors.

The D-Wave One used current biased SFQ \cite{likharev,bunyk2001rsfq}
demultiplexer circuitry to address all $N$ programmable devices on
chip, requiring an asymptotically optimal $O(\log(N))$ number of
control lines. This circuitry was designed with consideration for the
need to minimize static power dissipated on chip during programming,
and to this end used very low bias supply voltages, and very low value
shunt resistors.  The predicted peak temperature of the junction
shunts during programming, about $500\,\mathrm{mK}$, was sufficiently
low to ensure negligible thermally induced bit errors in DAC
programming.

We found, however, that it took of order $1\,\mathrm{s}$ for this heat
to dissipate sufficiently for the processor to return to
$\sim20\,\mathrm{mK}$, a temperature low enough to run the quantum
annealing algorithm and obtain solutions to posed problems with
appreciable probability. As typical computation time is $\sim 20
\mathrm{\mu s}$, this was clearly an unacceptable amount of time to
wait to run the algorithm after programming.

D-Wave Two control circuitry was designed to eliminate static power
dissipation completely and to simplify the design greatly with XYZ
addressing scheme requiring $O(\sqrt[3]N)$ lines. Though not
logarithmic, this scaling is sufficiently weak to allow processors
significantly larger than the one under discussion to be operated in
our existing apparatus.

% \footnote{We have also considered other
%   approaches which would achieve logaritmic (using binary trees), or
%   even constant (using shift registers) scaling of the number of lines
%   used to program the system, while eliminating static power
%   dissipation by employing quantum flux parametron (QFP) based logic,
%   but it was not required for our current or several generations into
%   the future processors.}

Improvement in performance of the annealing algorithm was also a
central design goal of the D-Wave Two. In a fixed temperature
environment, performance can be improved by increasing the qubit
energy scale. This can be done by decreasing qubit inductance and
capacitance. Given constraints on dielectric permittivity and wiring
geometry, this is most practically accomplished in our design by
reducing the physical length of the qubit wiring. Reduction of length
by a factor of two compared with D-Wave One was achieved by adding two
metal layers to the fabrication process, for a total of 6
superconducting metal layers. This allowed for an increase of overall
processor density by a factor of four.

In the first section of this paper we will step back and discuss the
requirements of the whole chip and give our rationale behind the
chosen hardware graph topology (common between D-Wave One and Two)
from the top-down perspective.  We will then be in a position to
present the chosen implementation of D-Wave Two control circuitry in
bottom-up fashion. Read-out infrastructure will be described in a
subsequent publication.

\section{Circuit Topology\label{sec:top}}

D-Wave quantum annealing processors have evolved through a series of
generations subject to the competing pressures exerted by
computational complexity and practical implementation. They are
designed to solve this problem: given {\em hardware graph} $G$,
minimize the following quadratic form over discrete variables $s_i \in
\{-1, +1\}$:
$$
E(\vec{\mathbf s}|\vec{\mathbf h}, \hat{\mathbf J})=\sum_{i\in
  nodes(G)} {h_i s_i} + \sum_{<i,j>\in edges(G)}{J_{ij} s_i s_j},
$$ with problem parameters $h_i,J_{ij}\in \{-1, -7/8, ... +7/8, +1\}$
for our current hardware.

Our current hardware graph topology, which we named Chimera, was
designed to satisfy a number of common-sense requirements related to
its intended use for solving optimization problems, subject to a
number of physical implementation constraints.

\subsection{Requirements}
\begin{enumerate}
\item{\bf Non-planarity:} We desire to tackle NP-complete problems, so
  non-planarity of the underlying graph is an important condition to
  making the corresponding Ising-spin problem NP-complete
  \cite{Barahona82,istrail2000statistical}. A related motivation is
  that non-planarity is required to establish chains of qubits that
  cross each other.

\item{\bf The ability to embed complete graphs:} To solve Ising spin
  glass problems with different topologies using a single processor, it
  must be possible to map, or {\em embed} the problem graph into the
  available hardware graph. This process typically involves using
  chains, trees, or other connected sub-components of the hardware
  graph (comprising {\em physical qubits}, strongly ferromagnetically
  coupled to each other) to represent a single node in the problem
  graph ({\em logical qubit}). In the language of graph theory, we
  want the hardware graph to have largest possible variety of problem
  graphs as its {\em minors} \cite {Vicky2008}. While embedding
  arbitrary graphs is computationally hard, one can nonetheless
  determine how large of a degree $M$ complete graph $K_M$ can be
  embedded in a given size hardware graph, thus guaranteeing that all
  graphs up to $M$ nodes are embeddable using a straightforward
  prescription.
   % \footnote{Note
   %  that we are not making any claims about how large a chain can we
   %  reliably maintain in a physical implementation of our processors,
   %  given realistic qubits having certain energy scales and subject to
   %  realistic levels of environmental noise.  Rather, we are only
   %  concerned herein with logical possibility of implementing such
   %  chains.}

\item{\bf The ability to incorporate on-chip control circuitry:} While
  a single qubit, or a handful of them, can be precisely controlled
  with dedicated analog lines driven by room-temperature electronics,
  integrating more than a few dozen qubits on a single chip requires
  some on-chip control circuitry. For example, there are currently six
  control ``knobs'' for every qubit, required to make the qubits robust
  to fabrication variability~\cite{CCJJqubit}, and one ``knob'' per
  coupler.  Where possible, we have designed our QA processors to use
  {\em static} flux biases applied to target superconducting loops in
  order to realize most of these knobs. The desired values of flux biases
  are programmed into individual control devices using a relatively
  small number of essentially digital control lines that carry signals
  generated at room temperature. These the control devices combine the
  functions of persistent memory and digital-to-analog conversion. We
  call these devices flux DACs, or \fdac s.  From an architectural
  viewpoint, each \fdac\ is a relatively macroscopic object with a
  typical size of $\sim 10\,\mathrm{\mu m}$.  Having several of them attached
  to a single qubit sets a lower bound on qubit size and influences
  possible qubit shapes and hardware graph topologies.
\end{enumerate}

\subsection{Constraints}
\begin{enumerate}
\item{\bf Limited qubit fan-out:} From an applications standpoint, the
  best option would be for each qubit to be connected to all others.
  However, directly implementing a complete $K_M$ graph in hardware
  for an arbitrarily large $M$ is impractical. Each qubit in our
  current design can be connected to only a relatively small number of
  other qubits $\lesssim 10$ before non-ideal features arise in the
  qubit response and the coupling energy scale (compared to $k_b T$,
  for example), becomes too small.

\item{\bf Minimizing uncoupled qubit/coupler lengths:} To optimize
  qubit energy scales and coupling strengths, neither qubits nor
  couplers can be designed to span an arbitrary length ({\it e.g.},
  full size of a large processor matrix). Ideally, all qubit length
  should be magnetically coupled to connected couplers, and all
  coupler length to connected qubits.

\item{\bf Minimization of noise pick-up areas and cross-talks:} Flux
  qubits and couplers are rf SQUIDs, which can be quite sensitive to
  magnetic fields.  Extreme care needs to be taken to minimize their
  pick-up of undesired disturbances, such as coupling to external flux
  noise sources or the unintended coupling ({\em cross-talk}) from a
  control line to a device it is not intended to control.

\item{\bf 2D chip integration:} While it would be nice to be able to
  grow a processor lattice in all three dimensions, in reality these
  lattices have to be implemented on the surface of 2D chips. Even if
  we imagine adding more metal layers to our fabrication process or 3D
  integration of several chips stacked on top of each other and
  passing quantum state between them (through, {\it e.g.},
  superconducting backside vias), growing the processor graph along
  the {\em physical} third dimension will always be harder than along
  the 2D chip plane.\footnote{Of course, one can attempt embedding
    {\em logically} higher-dimensional structures into planar chips,
    but this approach soon runs afoul of the second constraint above.}

\item{\bf Regularity and the notion of a ``unit tile'':} While it is
  in principle possible to arrange qubits in highly irregular
  structures, in practice, especially while designing chips not
  tailored to any specific problem graph structures which might arise
  in a concrete application, we find it convenient (to simplify the
  design and operation) to introduce the notion of a {\em unit tile},
  which is a smaller structure that can be replicated in both
  dimensions of the chip plane.
\end{enumerate}

\subsection{Chimera topology}

\begin{figure}[tb]
\includegraphics[width=\columnwidth]{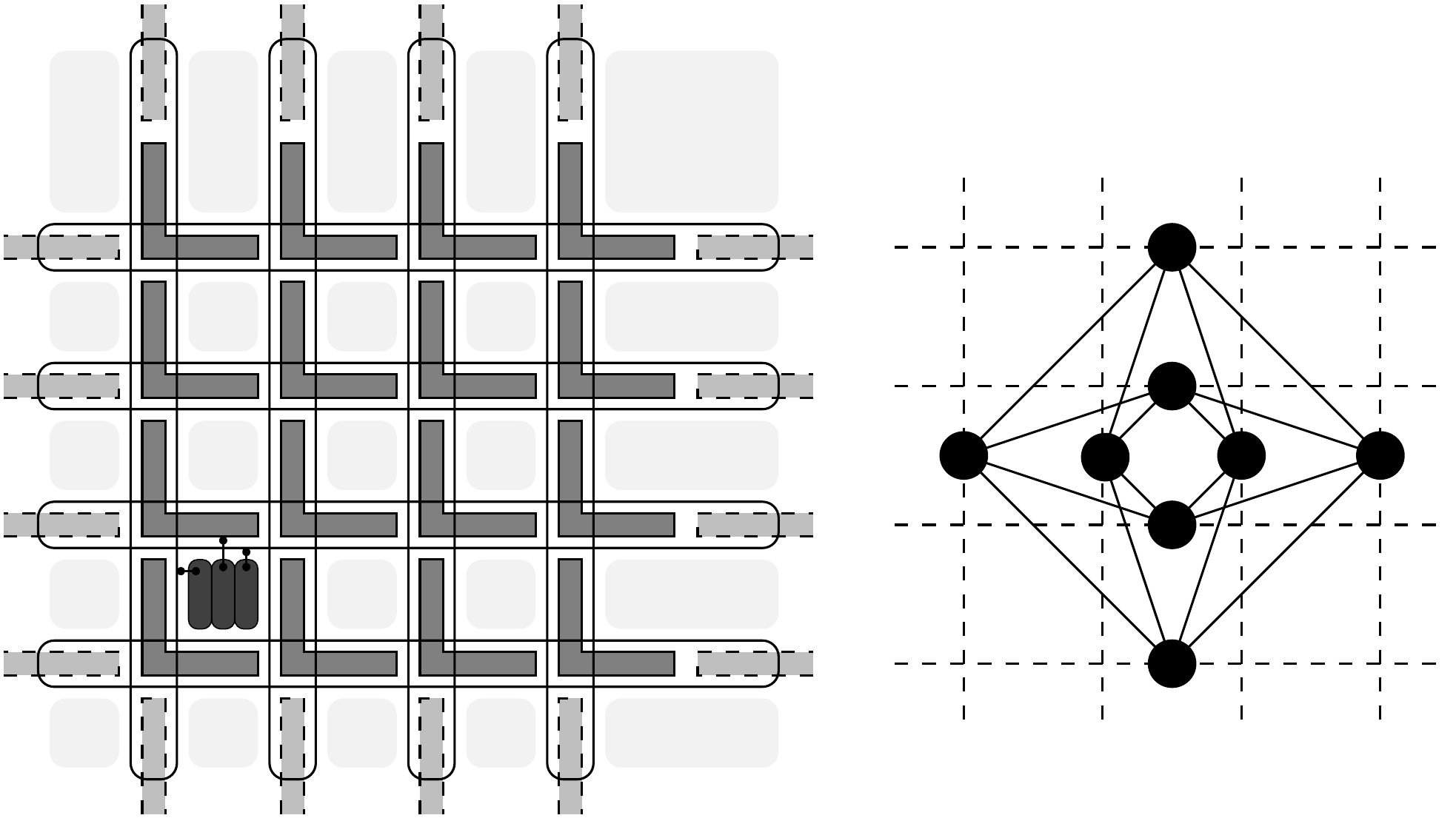}    
\caption{Chimera unit cell topology.  (Left) Layout sketch: qubit
  bodies are represented by the elongated loops that span the whole
  width/height of the unit tile. Each qubit is coupled to four others
  within the unit tile via the {\em internal coupler} bodies (dark L-shaped objects). Qubits are coupled to others in
  neighboring tiles via {\em external couplers} (lighter dashed rectangles). Control circuitry ({\fdac}s and
  corresponding analog control structures) are placed within
  light-shaded areas between the qubit/coupler bodies.  (Right) Graph
  representation: each unit tile corresponds to a complete bipartite
  graph $K_{4,4}$ (dark nodes and solid line edges).  Qubits from
  different tiles are coupled in square grid fashion (dashed edges).
\label{fig:chimeratop}}
\end{figure}

One of the features of a flux qubit is that (unlike, {\em e.g.,}
qubits based on quantum dots or individual trapped ions) they are
essentially macroscopic inductive loops interrupted by Josephson
junction(s) and that these qubit body loops can be stretched and
routed as needed. The same is true for qubit-to-qubit couplers
\cite{van2005mediated,harris2007sign,CJJcoupler}, except that
parametrically they tend to be lower inductance devices, and thus,
shorter.

With this in mind we examined different arrangements of qubit loops,
and eventually settled on the Chimera unit tile topology (used in both
D-Wave One and D-Wave Two processors), schematically depicted in
Fig.~\ref{fig:chimeratop}.  Each unit tile consists of 8 qubits -- 4
horizontal and 4 vertical -- with couplers between each
horizontal/vertical pair.  The unit tile is a complete bipartite graph
$K_{4,4}$. Unit tiles can be arranged into larger grid-like structures
that fill a plane, and each horizontal qubit can be coupled to the
corresponding qubits in the neighboring tiles to the left and right,
while each vertical qubit can be coupled to those in the neighboring
tiles above and below.

\begin{figure}[t]
\begin{center}
\includegraphics[width=0.75\columnwidth, angle=90]{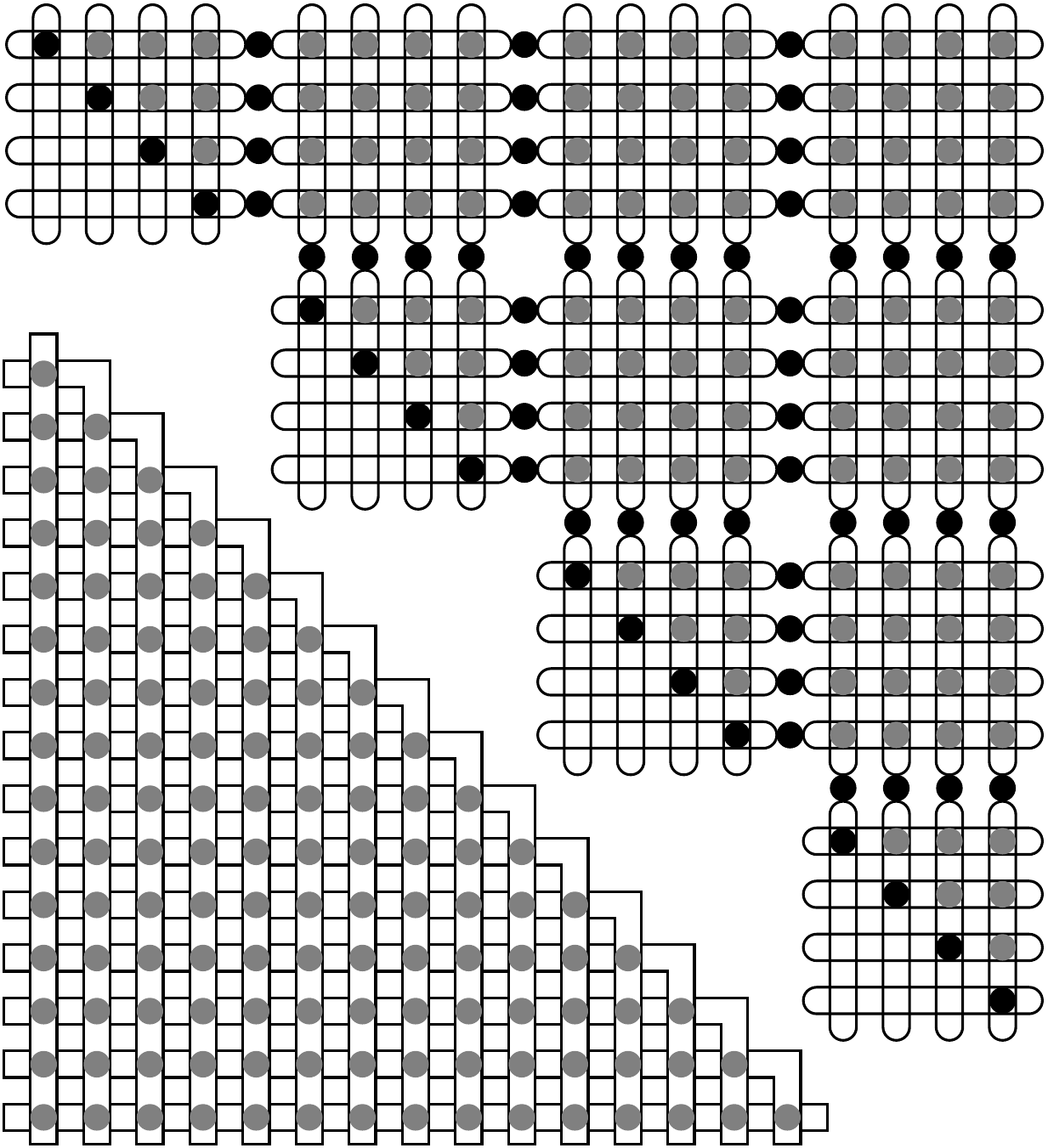}  
\end{center}  
\caption{Embedding complete graphs in the Chimera topology.  (Top)
  Strong ferromagnetic couplings (black circles) along the diagonal of
  the $C_N$ matrix ($N=4$ in this example), as well as along rows and
  columns between tiles in half of a $C_N$ (upper-left triangle shown)
  connects 80 physical qubits into 16 chains.  (Bottom) Each chain can
  be coupled to every other chain via sign and magnitude tunable
  couplers (gray circles), thus embedding a complete graph
  ($K_{4N=16}$ in this example). \label{fig:chimeraKN}}
\end{figure}

How well does the Chimera topology satisfy the requirements and
constraints given above? Consider the following:

\begin{itemize}
\item{The Chimera graph is non-planar.  Assuming the ability to
    establish chains of qubits along rows and columns of the processor
    matrix, there is a straightforward approach to embed complete
    graphs up to $4N$ nodes in a $N \times N$ grid of unit cells
    (denoted as $C_N$). This is illustrated in
    Fig.~\ref{fig:chimeraKN} for the case of $N=4$.  This approach can
    be validated based on the following observations:
\begin{enumerate}
\item Taking a single $K_{M,M}$ tile and ferromagnetically coupling
  pairs of horizontal and vertical qubits along its diagonal ({\em
    contracting edges}, which connect them in graph-theoretical
  language) produces a complete graph $K_M$.
\item Taking a $2 \times 2$ array of complete bipartite graphs
  $K_{M,M}$ and ferromagnetically coupling pairs of qubits in the same
  row/column produces a complete bipartite graph $K_{2M, 2M}$.
\item Taking two complete graphs $K_M$ and connecting them to two
  sides of a complete bipartite graph $K_{M,M}$ produces a complete
  graph $K_{2M}$ - every node in $K_{2M}$ can be coupled to every
  other node either because they either belong to the same $K_{M}$ and
  were coupled anyway or because they belong to different $K_{M}$s, in
  which case there is connection between them through the complete
  bipartite part.
\end{enumerate} }
\item{The Chimera topology was designed to be interleaved with the
    required control circuitry (which is schematically represented by
    lighter shaded areas in Figure \ref{fig:chimeratop}). In the
    implementation under discussion, each square ``plaquette'' formed at the
    intersection of two qubits contains three \fdac s. Generally, the left
    (right) \fdac\ provides certain type of control to the vertical
    (horizontal) qubit, while the middle one controls the
    corresponding coupler, as schematically shown in the bottom-left
    corner of this diagram.  }
\item{Almost all of the qubit length is coupled to couplers, and
  almost all of the coupler length is coupled to qubits, thus
  maximizing coupled signal strength. Also, implementing the qubit and
  coupler loops as long and narrow differential microstrip lines (in
  practice, over a superconducting ground-plane) minimizes noise and
  parasitic cross-talk pick-up. }
\item{Chimera unit tiles can be arranged into arbitrarily large 2D
  structures (limited only by fabrication yields, die size and
  available number of IO lines/die pads required to program all \fdac
  s). For example, the D-Wave One processor contained 128 qubits in a
  $C_4$ grid (a $4\times 4$ grid of 8-qubit tiles) and the D-Wave Two processor
  contains 512 qubits in a $C_8$ grid (an $8\times 8$ grid of 8-qubit tiles).}
\end{itemize} 

While this approach can be generalized to an arbitrary $K_{M,M}$ unit
tile with $2M$ qubits, our current implementation of $M=4$ was chosen
because (as will be seen later) all of its required \fdac s can be fit
in a $5 \times 5$ array of fixed-size plaquettes without too much
wasted space, simplifying (manual) layout and managing overall design
complexity. Another advantageous feature of the $M=4$ size is that the
number of \fdac s used for problem specification (one per qubit and
one per coupler, thus giving a total of 32 per tile) is approximately
balanced with the number of \fdac s used to make qubits robust against
fabrication variations (4 per qubit in the D-Wave One generation, 5 in
D-Wave Two, giving a total of 32 or 40 per tile, respectively). For
smaller unit tile size majority of the \fdac s would be of the second
variety.

\section{Design and operation of a \fdac.}

The precision desired for setting problem parameters sets the requirements for the range and
precision of individual \fdac s. Generally, our current implementation
requires about 8 bits of dynamic range for individual \fdac s, with
full ranges varying from several thousandths of a magnetic flux quantum
($\mathrm{m\Phi_0}$) to half a $\Phi_0$ coupled into qubit or coupler control
loops, depending on the \fdac\ type.

\begin{figure}[t]
\begin{center}
\includegraphics[width=0.5\columnwidth]{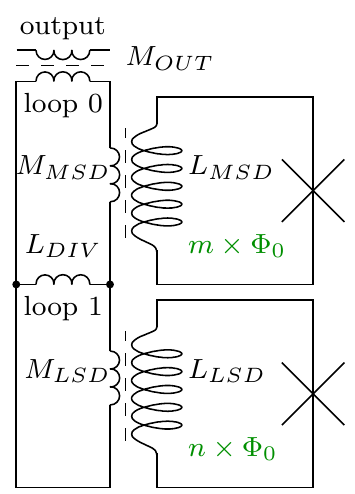}  
\end{center}  
\caption{Schematic view of generic two-stage D-Wave Two \fdac. 
\label{fig:DAC1}}
\end{figure}

To achieve this dynamic range while minimizing both total area occupied by
control circuitry (thus minimizing qubit length and increasing qubit energy
scales) and total number of wires needed for programming, in our
current design we chose to implement most of our \fdac s as two-stage
devices, of the kind schematically shown in Figure \ref{fig:DAC1}.

Each of the DAC digits (referred to as ``most significant'' and
``least significant'' here, or ``MSD'' and ``LSD'') is implemented as
a SQUID loop into which we can write and store some number of flux quanta $m$, {\it e.g.}, $-8 \lesssim m \lesssim 8$.
Individual quanta can be added to or subtracted from the storage loop
via  an SFQ pulse source, depicted here as a Josephson junction; its
structure and operation will be described in section
\ref{sec:vscrc}. Both digit storage loops are magnetically coupled
into an output device via an inductive ladder.

A single flux quantum added to the MSD coil induces
$M_{MSD}/L_{MSD}*\Phi_0$ flux into the top ladder loop, and
$M_{MSD}/L_{MSD}*M_{OUT}/L_{loop0}^{total}*\Phi_0$ into the output
device ($L_{loop0}^{total}$ denotes total inductance of the MSD
loop). The output flux increases proportinally with the number of flux
quanta added, up to a maximum determined by the device parameters and
addressing scheme. There is in general a nonlinear component
associated with the junction inductances, but as long as these
inductances are small compared to main loop inductances (true for our
devices), this correction is negligible. In our example, if MSD loop
can store up to 8 flux quanta of either polarity, it can provide 16
distinct values of output flux, or implement a 4-bit DAC.

To increase precision, a second stage is added. Here, the effect of a single
flux quantum in the LSD loop is further subdivided by a factor of
$L_{DIV}/L_{loop1}^{total}$. If this loop can also provide 16
distinct values of stored flux and the division ratio is 16, one
MSD step will be further subdivided into 16 steps of LSD, and the
two-stage device is an 8-bit DAC.

In practice, of course, we want to guarantee both the total output range
and the coverage of an MSD step by the LSD in the presence of fabrication
variations, so we need to add some margin to the number of quanta that
we can store in both loops.

\fdac s with different numbers of digits and weights of each digit can
be designed using the same principles, but we have found that this
two-stage design is sufficient for almost all of our DACs\footnote{Two
  special cases were introduced in D-Wave Two processors: a DAC which
  biases the qubit CCJJ major loop, for which currently 5 bits of
  dynamic range is sufficient and it was implemented as a single coil
  of the same type directly coupled into a target device, and second a
  very coarse stage for a qubit flux bias DAC, useful to deal with
  larger local qubit flux offsets.}.

\subsection{\fdac: Inductive storage and ladder}

\begin{figure}[t] 
\begin{center}
\includegraphics[width=\columnwidth]{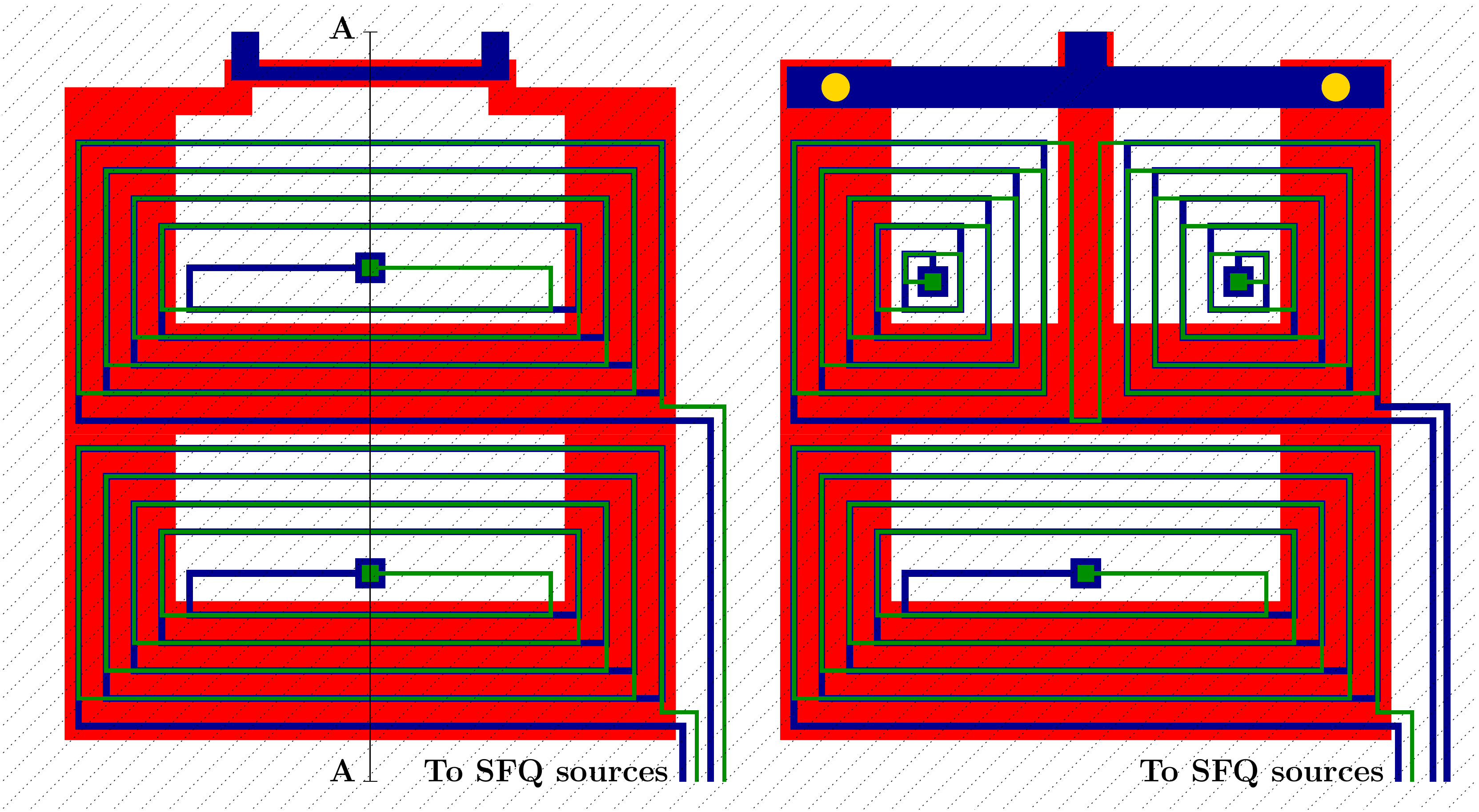}
\vskip3mm
\includegraphics[width=\columnwidth]{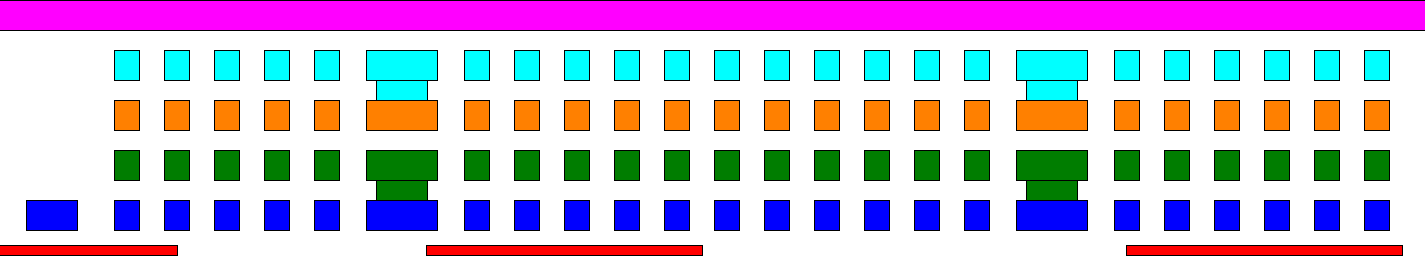} 
\color{black}
\end{center} 
\caption{Top: Simplified layout of two typical \fdac s. Low range,
  magnetically coupled to the output (left) and high-range ({\it
    e.g.}, coupler \fdac), merging device to be controlled (compound
  Josephson junction) with MSD ladder stage (right). Leads of the
  coils connect to SFQ pulse sources (not shown).  Bottom: CAD view of
  a cross-section of a real \fdac\ layout (to scale; width and spacing
  of spiral wires are $0.25\,\mathrm{\mu m}$) along cut line similar to line A-A
  shown on the low-range (left) device.}
\label{fig:DAC2}
\end{figure}

Having covered the basic idea behind our \fdac s, we can present a more
realistic layout of their implementation on the top of Figure
\ref{fig:DAC2}. Large storage inductors ($\sim 1\, \mathrm{nH}$) are
implemented as stacked spirals (blue and green in the figure),
shown here wound in two metal layers, though our real layouts use four
layer spirals, with $0.25\,\mathrm{\mu m}$ line width and spacing design
rules. 

The inductive ladder is implemented as two galvanically connected washers
in the bottom metal layer (red), magnetically coupled to their two
coils. The horizontal bar between them implements the shared inductance
$L_{DIV}$ of Figure \ref{fig:DAC1}.

To minimize unintended coupling between DAC coils and other elements
of the circuit, the whole structure is covered by a shielding
sky-plane in the top metal layer (dotted diagonal lines).

\begin{table*}
\begin{center}
\caption{\fdac\ parameters calculation \label{tab:fdac}}    
\setlength{\extrarowheight}{8pt}
\begin{tabular}{r|rcl}
  Weight of one SFQ in LSD loop, $\mathrm{m\Phi_0/\Phi_0}$ & $W_{LSD}$ & = & $
  1000*\left(\frac{M_{LSD,OUT}}{L_{LSD}} - \frac{M_{LSD,MSD}}{L_{LSD}}*\frac{M_{MSD,OUT}}{L_{MSD}}\right)
  $ \\
  Weight of one SFQ in MSD loop, $\mathrm{m\Phi_0/\Phi_0}$ & $W_{MSD}$ & = & $
  1000*\frac{M_{MSD,OUT}}{L_{MSD}}
  $ \\ 
  LSD capacity, $\Phi_0$ & $\mathrm{MAXSFQ}_{LSD}$ & = & $\left \lfloor
  {I_{in}*L_{LSD}/\Phi_0} \right \rfloor$ \\
  MSD capacity, $\Phi_0$ & $\mathrm{MAXSFQ}_{MSD}$ & = & $\left \lfloor
  {I_{in}*L_{MSD}/\Phi_0} \right \rfloor$ \\  
  Division ratio, should be $\leq \mathrm{MAXSFQ}_{LSD}$ & & & $W_{MSD}/W_{LSD}$ \\
  Total range, $\mathrm{m\Phi_0}$ & $Range$ & = & $W_{MSD}*\mathrm{MAXSFQ}_{MSD}$ \\
  Effective dynamic range, bits & & & $\log_2\left|\frac{Range}{W_{LSD}}\right|$ \\
\end{tabular}
\end{center}  
\end{table*}

Simple magnetic coupling between the inductive ladder and target
device using the microstrip transformer shown in the top-left panel of
Figure \ref{fig:DAC2} is sufficient to implement a full range of several
tens of $\mathrm{m\Phi_0}$ into the target device. However, the majority of our
DACs (ones that control the compound Josephson junctions of couplers,
inductance tuners and persistent current compensators) require a range
comparable to half a $\Phi_0$, since they need to be able to bias their target's
corresponding CJJ all the way between its maximum $I_c$ and fully
suppressed. To implement such higher-range control we merge the CJJ loop
of the target device with the MSD stage of the inductive ladder, as shown in
the top-right panel of Figure \ref{fig:DAC2} (Josephson junctions
are shown as yellow circles).

An additional complication of this particular structure is that the
DAC should be coupled with equal strength into both halves of the
target CJJ loop in order to avoid coupling into target device body. To
achieve this, the MSD coil is split into two symmetric halves, as
shown in the top-right panel of Figure \ref{fig:DAC2}.

The simplistic lumped-element model of Figure \ref{fig:DAC1} is not
entirely adequate as a complete description of our \fdac\ devices,
especially considering the cross-section of an actual device
implemented in all six available metal layers (drawn to scale) at the
bottom of Figure \ref{fig:DAC2}. The LSD loop couples flux directly
into the MSD one and it can reach the output not only via the
inductive ladder, but also via this magnetic connection to the
(strongely coupled to the output) MSD. In addition, the MSD flux can
reach the output directly, not mediated by the washer (and with the
sign opposite to the washer-mediated coupling).

We treat a complete \fdac\ structure as a three-port device (LSD, MSD
and OUT) and, using the 3D inductance extraction program {\tt
  FastHenry} with superconductor support \cite{wrcad}, extract its
complete inductance matrix:
$$
\left(
\begin{matrix}
L_{LSD} & M_{LSD, MSD} & M_{LSD, OUT} \\
M_{LSD, MSD} & L_{MSD} & M_{MSD, OUT} \\
M_{LSD, OUT} & M_{MSD, OUT} & L_{OUT}   
\end{matrix}
\right).
$$

For subsequent analysis we treat the SFQ pulse sources as simple
current sources that can produce (up to) $I_{in}$ (approximately half
of their junction critical current $I_c$, as discussed below) into a
large inductive load, and calculate all relevant parameters of our
\fdac s as shown in Table \ref{tab:fdac}. After we build a \fdac\
layout model, we iterate over its geometrical parameters to ensure
that it fits into the available space, has the required number of bits
and range, and that its MSD/LSD division ratio is such that the LSD
comfortably spans a single MSD step.

\subsection{\fdac: SFQ pulse sources\label{sec:vscrc}}

Our implementation of an SFQ source is based on perhaps the earliest
incarnations of single-flux-quanta circuits
\cite{silverHurrellPridmorebrown}: a current biased dc-SQUID made
with two shunted junctions.

\begin{figure}[t]
\begin{center}
\includegraphics[width=0.9\columnwidth]{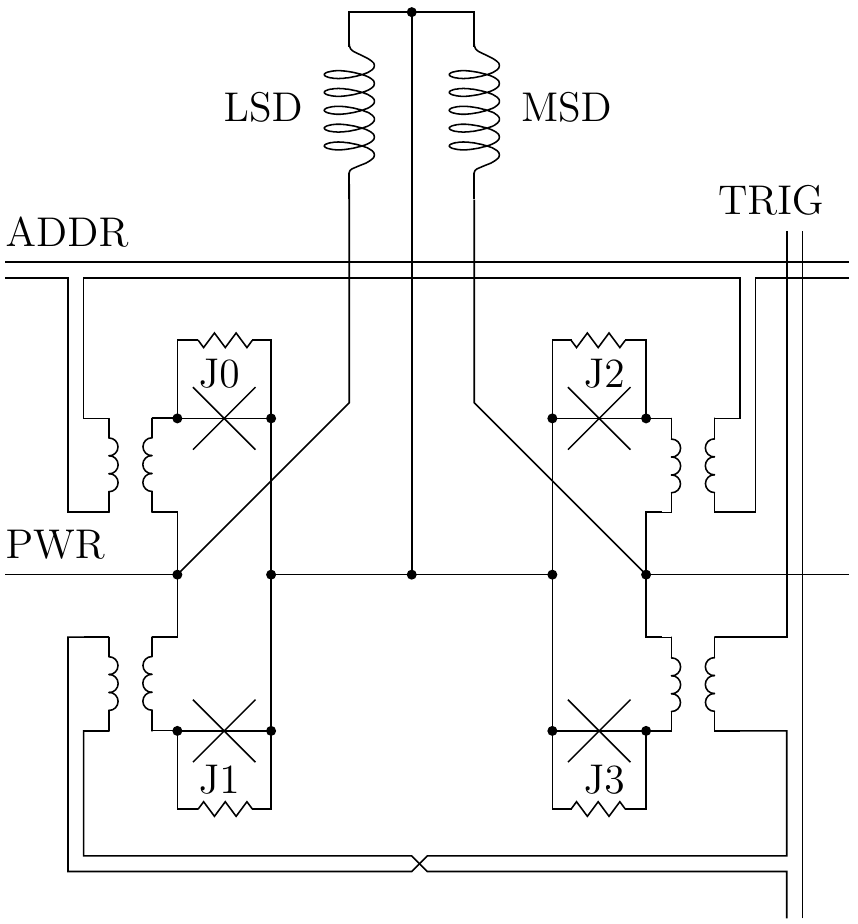}
\end{center}  
\caption{Schematic of two SFQ sources feeding two stages of a \fdac.
  There are 4608 such pairs implemented on a 512 qubit processor.
  Junction critical current is $55\mu A$.  Each junction is shunted
  with approximately 0.58 $\Omega$, which corresponds to a $\beta_c
  \simeq 0.05 $.  The DAC storage inductances for LSD and MSD loops
  are $ \simeq 1~nH$, whereas inductance of the source itself is
  $24~pH$.
%% % Notes on beta_c: Junctions are 8um x 2.75um = 22um^2 in area
%% % C ~ 40 fF/um^2 * 22um^2 = 0.88 pF
%% % betac = ( R C )/( L / R) = RC/(Phi0/(2 PI Ic R)) = 2 PI Ic R^2 C/Phi_0
%% % R is marked as 0.577 on the layout, and this works out to about 0.05 for betac
  \label{fig:Vsrc}}
\end{figure}

A schematic of two dc-SQUIID SFQ pulse sources feeding the LSD and MSD
storage loops of a single \fdac\ is shown in Figure \ref{fig:Vsrc}.

To operate, one first applies PWR current bias (biasing all junctions
to about half of their $I_c$). ADDR is then applied, providing an
initial flux bias to the dc-SQUID bodies. Ramping TRIG with a polarity
that adds to ADDR in, for example, the dc-SQUID comprised of junctions
J0 and J1, eventually steers enough current through J0 to exceed its
critical current, causing it to ``flip'' by $2\pi$ in phase, admitting
a single flux quantum into the dc-SQUID loop. TRIG is then decreased,
eventually causing J1 to flip. The J0/J1 dc-SQUID is thus returned to
its zero flux state, but in the process the phase drop across the LSD
inductor has been increased by $2\pi$ -- an SFQ pulse is added to that
storage loop. Assuming the LSD inductor is large compared to the
dc-SQUID inductance, this process can be repeated until the persistent
current stored in the LSD loop becomes comparable to the PWR current,
cancelling it, preventing junctions from further flipping. At that
point, the \fdac\ loop has reached its maximum SFQ capacity.

If one changes the sign of PWR, using the same process one can add
single flux quanta of the {\em opposite} magnetic field direction into
this storage loop (or, {\em subtract} from the ones stored there).

Note that the TRIG line is twisted between the dc-SQUIDs J0/J1 and J2/J3, so when
it adds to the ADDR prebias for the J0/J1 dc-SQUID, it subtracts from the J2/J3 dc-SQUID,
and the J2/J3 SQUID is quiescent. But if one reverses the polarity of
the TRIG pulses relative to the ADDR pre-bias, one can operate the J2/J3 dc-SQUID,
adding a SFQ to the MSD \fdac\ coil. The relative polarity of ADDR and TRIG
allows us to select the \fdac\ stage on which we want to
operate. 

The PWR, ADDR, and TRIG levels are chosen to meet the following
criteria:
\begin{enumerate}
\item{} With PWR held at its active level, the state of a \fdac\ changes
  by exactly one flux quanta per ADDR, TRIG pulse.
\item{} Each \fdac\ undergoes SFQ transitions only when all three lines
  addressing that device are active. If two or fewer lines are active,
  the state of the \fdac\ does not change.
\end{enumerate}
If these criteria are met, then a limited number ($O(\sqrt[3]N)$) of
control lines can address $N$ \fdac s in what we call ``XYZ'' fashion,
discussed further in Section \ref{sec:XYZ}. Here we discuss the
process, which we refer to as margining, by which programming levels
are chosen to meet the above criteria.

\fdac\ state stability is fully determined by $\Phi_b$ and $I_b$, where
$\Phi_b$ is the sum of the ADDR and TRIG flux biases and $I_b$ is the
total current biasing the dc-SQUID SFQ pulse source. $I_b$ includes
contributions from PWR and from the current circulating in the main
\fdac\ loop due to its flux state. A critical line in $(\Phi_b, I_b)$
space, similar to that of a current biased dc-SQUID, bounds the region
in which a flux state is stable. When this line is crossed due to
manipulation of PWR, ADDR, and TRIG, a transition will take place.

\begin{figure}[t]
  \includegraphics[width=8cm]{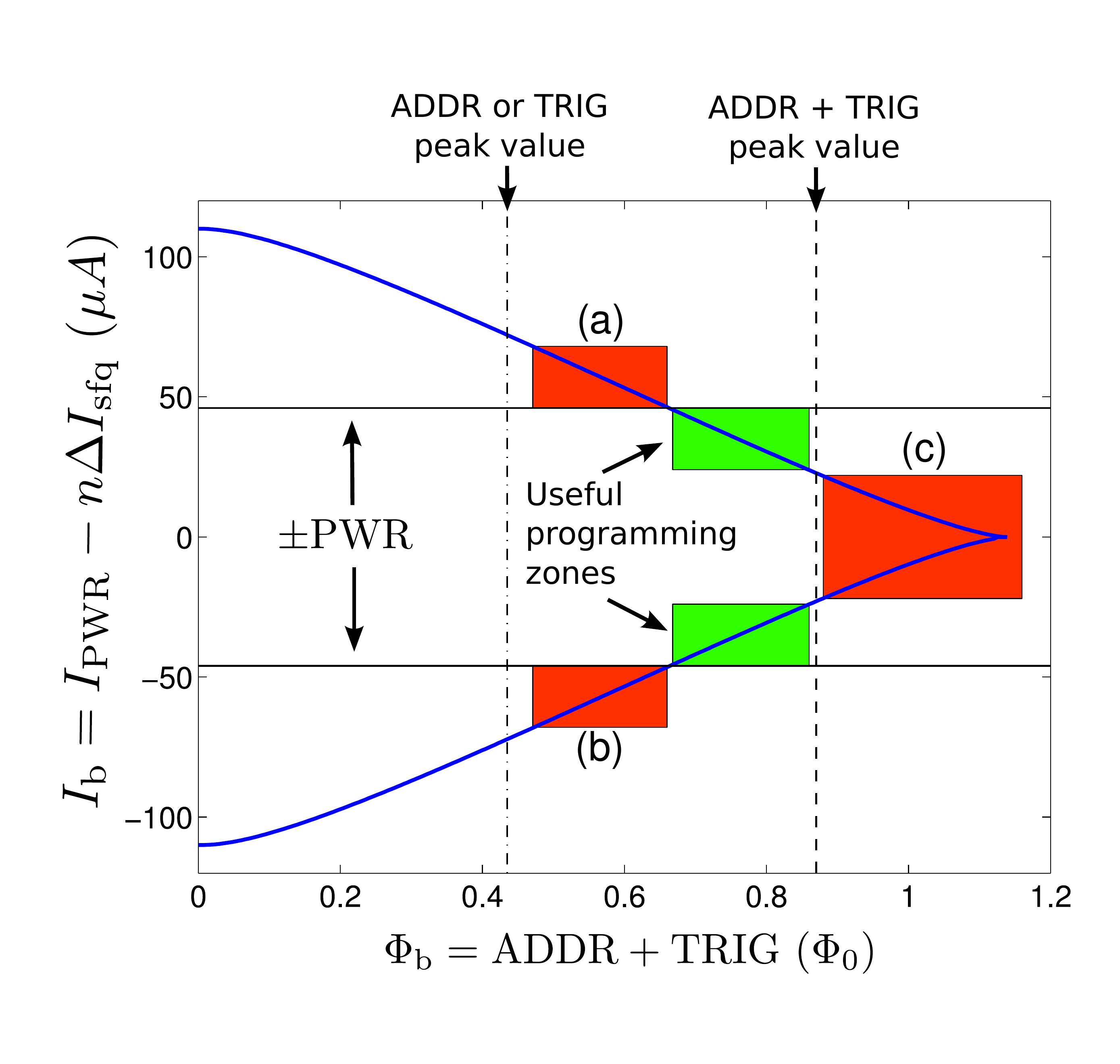}
  \caption{\fdac\ margining diagram. The position of a given \fdac\ on
    this diagram depends on its SFQ state and bias levels. The ADDR
    and TRIG levels determine the $x$-axis position. The PWR level,
    along with the flux state of the main loop of the \fdac{},
    determine the $y$-axis position. The blue curve is the critical
    line of the dc-SQUID pulse source zero flux state. A \fdac\ will
    undergo the first of the two junction flips necessary to increment
    the \fdac\ state when it crosses this line from left to right. The
    active PWR levels for programming +SFQ and -SFQ are the $\pm 45
    \mu\mathrm{A}$ horizontal black lines. Before adding $\pm N$
    pulses to a \fdac{}, the \fdac\ state is first RESET to zero. The
    green zones encompass the sections of the critical line where
    programming takes place when all three lines addressing a \fdac\
    are active. The red zones encompass values of $I_b$ that can exist
    in \fdac{}s addressed by a subset of the three line types. To
    avoid unwanted programming, $\Phi_b$ for those DACs must not enter
    these regions. \fdac{}s addressed by only PWR and either ADDR or
    TRIG can have $I_b$ values in regions (a) and (b), and therefore
    ADDR or TRIG levels must not reach these regions. \fdac{}s
    addressed by only ADDR and TRIG will have $I_b$ values in region
    (c), and therefore the sum of the ADDR and TRIG levels must not
    reach this region. The height of region (c) is equal to the
    combined heights of the two green regions, is equal to the
    combined heights of the outer red regions, and is equal to the
    main \fdac\ loop current range $I_{in}$. The critical line shown was
    computed using the average \fdac\ parameters measured on a D-Wave
    Two processor. The ADDR and TRIG levels chosen for this processor
    (vertical dashed lines) do not quite match the boundaries of the
    allowed and forbidden zones (as would be optimal) due to variation
    of physical parameters between \fdac{}s and the requirement that
    the margining criteria be met for all \fdac{}s.}
  \label{fig:margining}
\end{figure}

In Figure \ref{fig:margining}, the critical line of the zero flux
state of the dc-SQUID pulse source is plotted. Crossing this boundary
corresponds to the first junction flip in the SFQ pulse sequence
described previously. The system will cross the boundary at a point
that depends on the main \fdac\ loop flux state. Margining of the PWR,
ADDR, and TRIG levels can be understood as a geometric partitioning of
this boundary into active regions, in which intended transitions will
take place, and forbidden zones, in which transitions that do not meet
the margining criteria would occur. PWR, ADDR and TRIG levels are
chosen to maximize the size of the active regions while avoiding the
forbidden zones.

\subsection{\fdac\ reset}

The protocols described above allow us to add or subtract SFQ to a
\fdac\ stage, ending up with a known number of flux quanta when we start
from a known state. For realistic operation we must also be able to
reliably reset all \fdac s into a known state starting from an
{\em unknown} state.

To reset a \fdac, $I_{PWR}$ is set to zero. The dc-SQUID pulse source
still sees a non-zero current bias, as long as the \fdac\ is in a
non-zero flux state. Programming the \fdac\ under these conditions
will cause SFQ changes in the main \fdac\ loop that decrease this
current bias. Applying ADDR+TRIG pulses with large enough amplitude to
reliably drive transitions (larger than the maximum $\Phi_b$ value of
the critical boundary in Figure \ref{fig:margining}) will 'de-program'
the \fdac\, one SFQ at a time, until it reaches its lowest energy zero
SFQ state for which the circulating current is zero. To reliably reach
this zero SFQ state, junction critical current asymmetry must be
small: the critical current difference between the two junctions in
the dc-SQUID pulse source should be well under $\Phi_0/L$, where $L$
is the main loop inductance.

Note that the margining criteria are violated during reset. All
\fdac{s} are reset simultaneously.

\subsection{Minimizing \fdac\ footprint}

\begin{figure}[t]
\begin{center}
\includegraphics[width=\columnwidth]{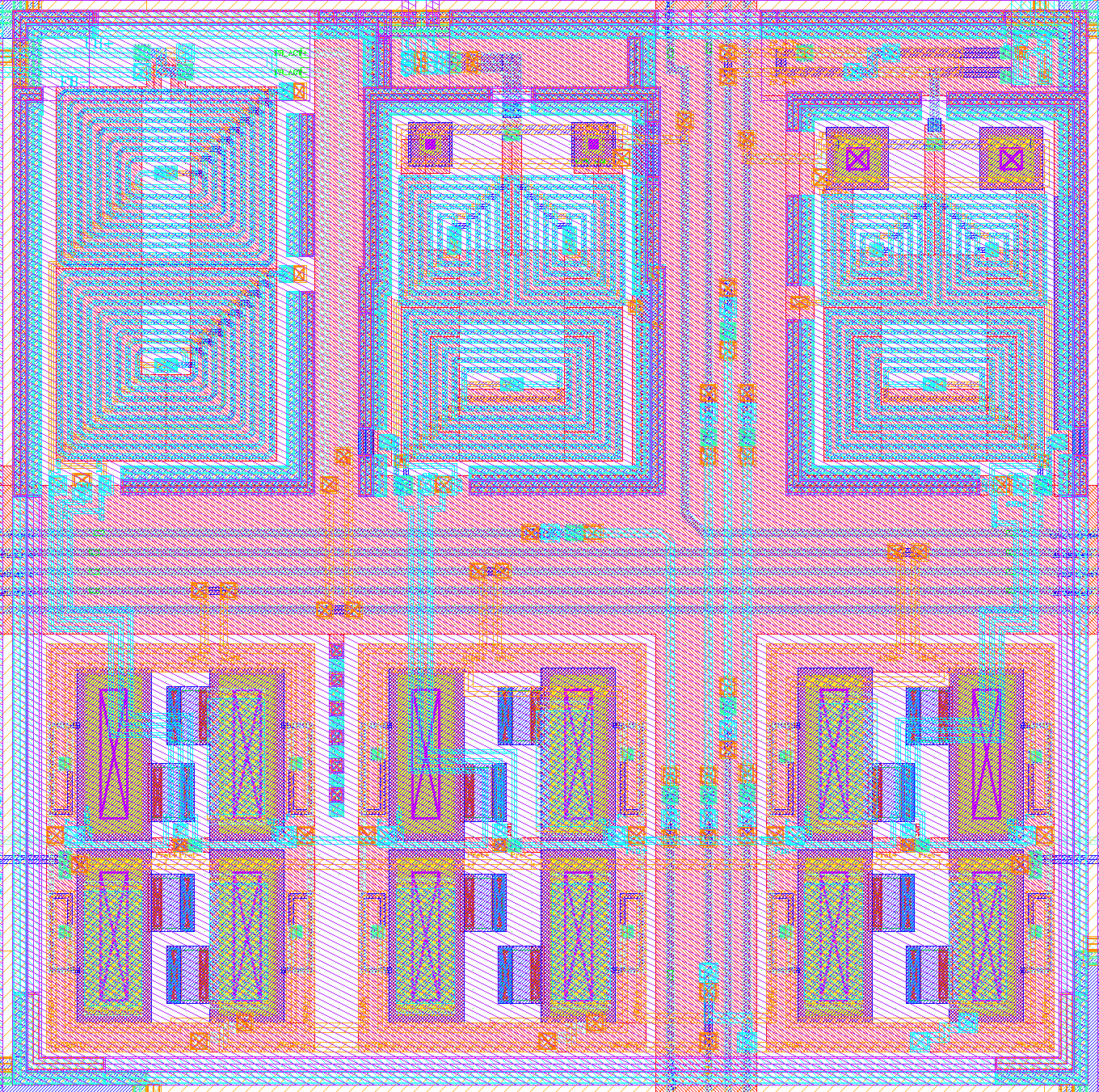}
\end{center}  
\caption{CAD layout of a single $60\times 60\, \mathrm{\mu m}^2$
  plaquette of our second generation 
  processor with three \fdac s. Note that areas of \fdac\ storage
  coils (top) and pulse sources (bottom, junctions are yellow
  rectangles) are approximately equal. 
  \label{fig:3DACs}}
\end{figure}

As we mentioned in Section \ref{sec:top}, the \fdac\ area is what
ultimately sets the size of our processor unit tile. This in turn
determined the length of the qubits, and thus their energy scales,
ultimately affecting the performance of the annealing
algorithm. Minimizing this area is therefore of great importance to
us.

What matters for a given \fdac\ to achieve its design objectives is
the maximum number of single flux quanta that we can store in its MSD
and LSD loops (determining maximum range and precision, provided that
division ratio is chosen correctly). That, in turn, is proportional to
the $L \times I_c$ product of the storage loop inductance $L$ and the pulse
source junction $I_c$. How can we minimize an area required to
implement both junctions and inductor to achieve constant (and
sufficient) $L \times I_c$ product? Equivalently, how can we maximize this
product in given area?

One can observe that (to the first order of approximation), inductance
of a spiral coil (for fixed number of available layers) is
proportional to its area with some proportionality constant
$\alpha$. The same is true for junction area, given a fixed critical
current density $J_c$. If we have unit area available for the whole
\fdac, and inductance occupies some fraction $x$ of that, $L \times I_c =
\alpha\, x \times J_c (1-x) \sim x \times (1-x)$, which reaches its maximum at
$x=0.5$, meaning that half of the optimal \fdac\ area is occupied by
storage inductance and another half by source junctions.

This was the rule that we used for choosing source junction $I_c$
vs.\ storage $L$ (their $J_c$ was fixed by requirements of the analog
qubit and coupler circuitry in a process with only a single available
trilayer). Figure \ref{fig:3DACs} is a CAD view of three \fdac s
within one plaquette of our current processor.

Note that the result of this analysis is independent of critical
current density. Suppose a second high-$J_c$ trilayer becomes
available for our next generation design, say, $9\,\mathrm{kA/cm^2}$
(in addition to our current $250\,\mathrm{A/cm^2}$), a factor of 36 in
$J_c$.  Just replacing the existing junctions with smaller in size and
equal critical current would save us less than a factor of 2 in \fdac\
area. If instead $L$ is decreased and $I_c$ is increased by a factor
of 6 in value, the total area decreases by the same factor, with
$L\times I_c$ product unchanged.

 \subsection{XYZ-addressing line count\label{sec:XYZ}}

\begin{figure}[t]
\begin{center}
\includegraphics[width=\columnwidth]{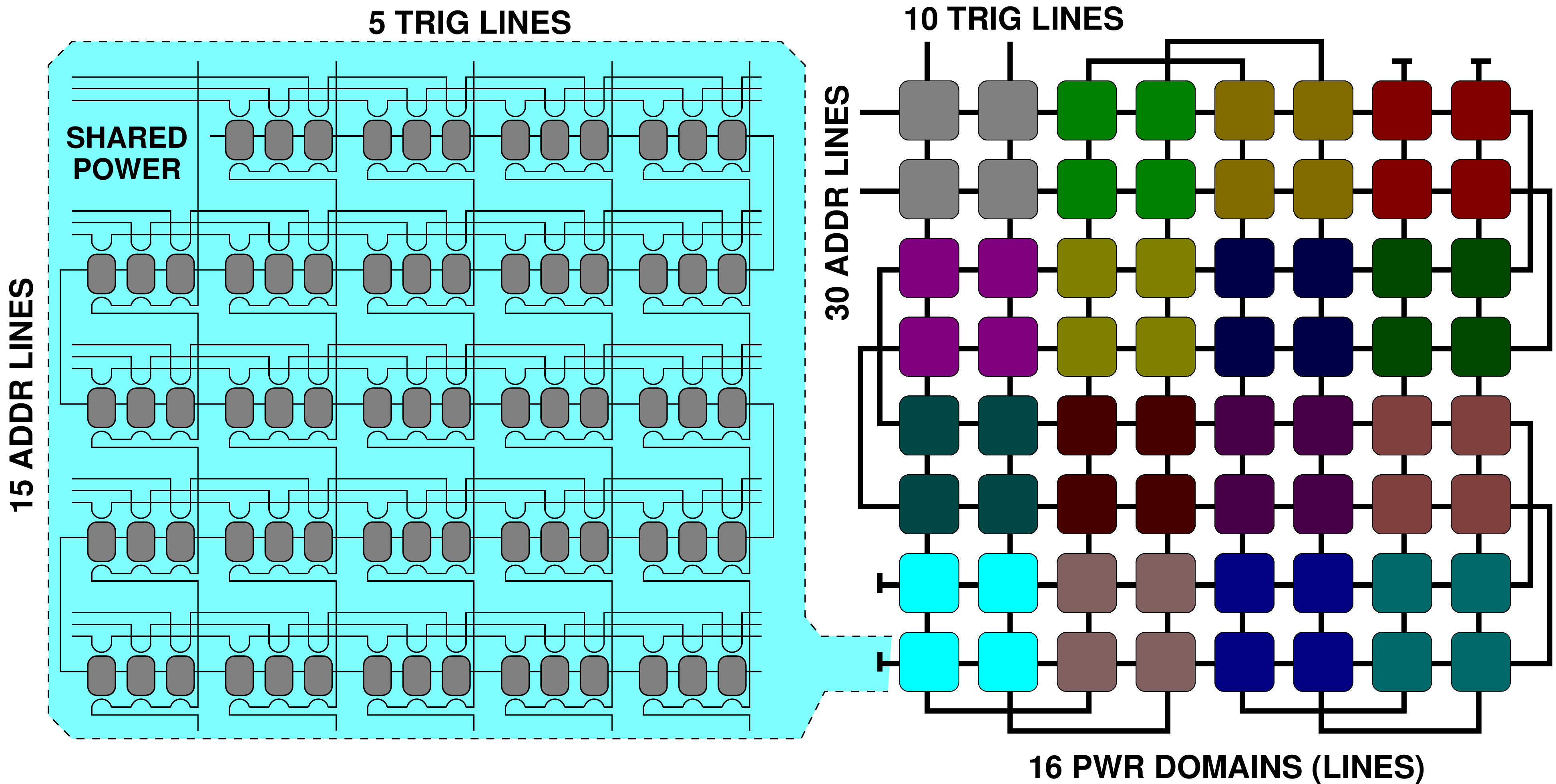}
\end{center}  
\caption{XYZ-addressing of \fdac s in the D-Wave Two processor. 72 \fdac s
  within a unit tile are selected using 15 ADDR and 5 TRIG lines
  (left). An $8\times 8$ array of unit tiles is split into 16 PWR
  domains (right), and all \fdac s within that can be addressed using
  30 ADDR, 10 TRIG and 16 PWR lines.
\label{fig:XYZ}}
\end{figure}

We need 72 \fdac s to control all the qubits and couplers of a unit tile
(6 per qubit, 16 for controlling internal couplers, and 8 for controlling external
couplers), for a total of 4608 \fdac s for our D-Wave Two 512-qubit processors. To
select one of them using cubic XYZ-addressing, we need at least $\left
  \lceil 3*\sqrt[3]{4608} \,\right \rceil = 50$ lines, or about 16 lines
per dimension.

We have arranged all required \fdac s for a given tile in 25 3-DAC
plaquettes (one plaquette is empty), as shown in the left panel of
Figure \ref{fig:XYZ}. One of three \fdac s within a plaquette is
selected using one of three ADDR lines, with all three sharing a TRIG
line, resulting in 15 ADDR and 5 TRIG lines addressing all \fdac s
within unit tile.

The third dimension of addressing is established by separating tile
arrays into PWR domains. Our D-Wave Two processors contain an $8\times
8$ array of unit tiles, split into sixteen $2 \times 2$ power domains,
as shown in the right panel of Figure \ref{fig:XYZ}.  All \fdac s
within one power domain are connected in series and fed by one of 16
PWR lines. 30 ADDR and 10 TRIG lines are reused between power domains,
for a total of $30+10+16=56$ lines used to address all \fdac s within
a processor matrix. While it is not optimal (because of the difference
in the number of ADDR and TRIG lines), it is sufficiently close, and
this arrangement allowed us to achieve a more regular layout without
having to assign different roles to a single line within the processor
fabric ({\em e.g.}, make a single line work as an ADDR for one DAC and
a TRIG for another).

\begin{figure}[t]
\begin{center}
\includegraphics[width=\columnwidth]{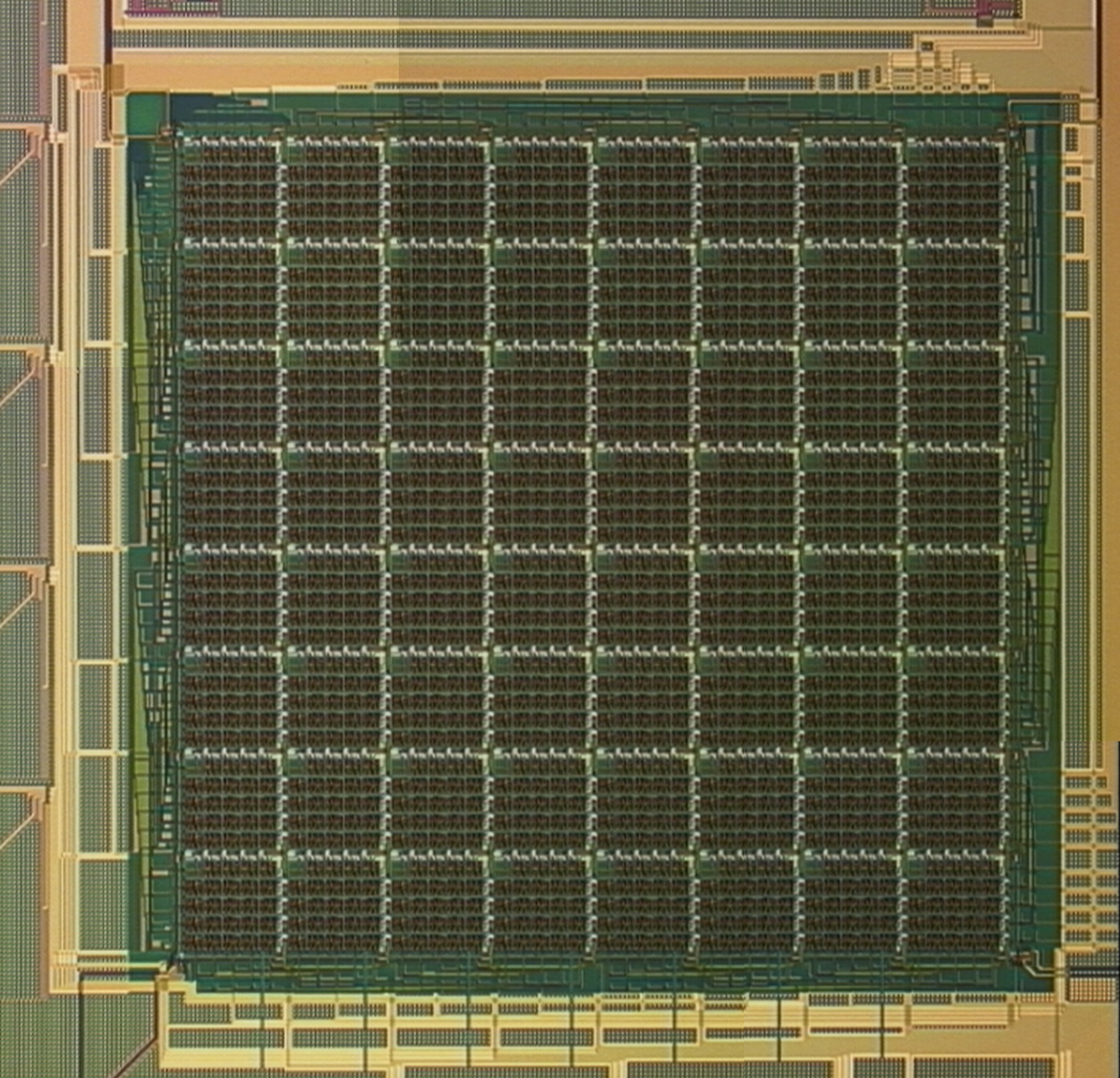}
\end{center}  
\caption{Microphotograph of an active portion ($\sim 3.5 \times 3.5
  \,\mathrm{mm^2}$) of D-Wave Two processor chip, $8 \times 8$
  array of 8-qubit unit tiles, one unit tile is $335\,\mathrm{\mu m}$ on the
  side. This picture was taken before deposition of the last metal
  layer (serving as skyplane), making internal structure visible.  
  \label{fig:photo}}
\end{figure}

\section{Conclusions}

We have described how, starting with top-level requirements of a
processor implementing a quantum annealing algorithm, we have designed
its hardware graph and required control infrastructure, which allowed
us to successfully operate processors with up to 512 rf-SQUID qubits
using only 56 control lines for problem programming. Figure
\ref{fig:photo} shows a microphotograph of an active area of a D-Wave
Two processor chip.

The most important feature of our new \fdac{} design is its zero
static power dissipation -- unlike traditional SFQ circuitry, which
incorporates on-chip resistive current sources tapping a common
voltage rail, this design biases all devices serially with a fixed
current whose magnitude is set by a {\em room-temperature}
resistor\footnote{This approach can also be viewed as implementing SFQ
  ``current recycling'' taken to its ultimate limit.}. The only energy
dissipated on-chip is on the order of $I_c \times \Phi_0$ per flux
quantum moved into (or out of) the storage inductor. For a pair of
$55\,\mathrm{\mu A}$ \fdac\ junctions this corresponds to
$0.22\,\mathrm{aJ}$. Complete reprogramming of all 9216 \fdac\ stages
moving from -16 to +16 SFQ in their storage loops would dissipate on
chip only about $65\,\mathrm{fJ}$.

While the D-Wave One required a post-programming delay of about $1\,
\mathrm{s}$, D-Wave Two can thermalise to $20\, \mathrm{mK}$ within
$10\, \mathrm{ms}$, a factor of 100 improvement achieved within one
processor generation just in this post-programming thermalization
time.

%\clearpage

\bibliography{master,2013-QA-processor-white-paper}
\bibliographystyle{IEEEtran}

\end{document}